\documentclass[conference,twocolumn]{IEEEtran}

\usepackage{enumerate}
\usepackage{algorithm,amsmath}
\usepackage{algpseudocode}
\usepackage{amssymb}
\usepackage{color}
\usepackage{xcolor}
\usepackage{graphicx}
\usepackage{amsmath}
\usepackage{subfigure}
\usepackage{multirow}
\usepackage{hyperref}
\usepackage{tabularx}

\usepackage{svg}

\usepackage{comment}

\usepackage{lipsum,mathtools}
\usepackage{bm}
\DeclarePairedDelimiter{\ceil}{\lceil}{\rceil}

\newcommand\blfootnote[1]{%
	\begingroup
	\renewcommand\thefootnote{}\footnote{#1}%
	\addtocounter{footnote}{-1}%
	\endgroup
}

\newcommand{\fr}[1]{\mathbf{\tilde{#1}}} 

\usepackage[compact]{titlesec}
\titlespacing{\section}{0pt}{2ex}{1ex}
\titlespacing{\subsection}{0pt}{1ex}{0ex}
\titlespacing{\subsubsection}{0pt}{0.5ex}{0ex}

\begin{document}
	
	\title{Optimization of Quantized Phase Shifts for Reconfigurable Smart Surfaces Assisted Communications}

	\author{\IEEEauthorblockN{Authors}
	}
	
	\author{Maximiliano Rivera\IEEEauthorrefmark{1}, Mohammad Chegini\IEEEauthorrefmark{2},
		Wael Jaafar\IEEEauthorrefmark{3}, \IEEEauthorblockN{Safwan Alfattani\IEEEauthorrefmark{4}\IEEEauthorrefmark{5}, Halim Yanikomeroglu\IEEEauthorrefmark{3}}
		\IEEEauthorblockA{\IEEEauthorrefmark{1}Pontificia Universidad Cat\'olica, Santiago, Chile, \IEEEauthorrefmark{2}Shahid Beheshti University, Tehran, Iran}  
		\IEEEauthorrefmark{3} Carleton University, Ottawa, ON, Canada, \IEEEauthorrefmark{4} University of Ottawa, ON, Canada
		
		\IEEEauthorrefmark{5} King AbdulAziz University, Saudi Arabia
	}
	
	
	\maketitle
	
	\begin{abstract}
		Reconfigurable Smart Surface (RSS) is assumed to be a key enabler for future wireless communication systems due to its ability to control the wireless propagation environment and, thus, enhance communications quality. Although optimal and continuous phase-shift configuration can be analytically obtained, practical RSS systems are prone to both channel estimation errors, discrete control, and curse of dimensionality. This leads to relaying on a finite number of phase-shift configurations that is expected to degrade the system's performances. In this paper, we tackle the problem of quantized RSS phase-shift configuration, aiming to maximize the data rate of an orthogonal frequency division multiplexing (OFDM) point-to-point RSS-assisted communication.   
		Due to the complexity 
		of optimally solving the formulated problem, we propose here a sub-optimal greedy algorithm to solve it. 
		Simulation results illustrate the 
		performance superiority of the proposed algorithm compared to baseline approaches. Finally, the impact of several parameters, e.g., quantization resolution and RSS placement, is investigated.
		
	\end{abstract}
	
	
	
	\vspace{-10pt}
	
	\blfootnote{This work is supported by Huawei Canada, the Natural Sciences and Engineering Research Council Canada (NSERC), and a scholarship from King AbdulAziz University,  Saudi Arabia.}
	
	\section{Introduction}
	
	
	{Reconfigurable smart surfaces (RSS) arise as a green and
		sustainable technology for beyond fifth generation (B5G) wireless networks, owing to their potential to enable cost-effective massive connectivity \cite{Alfattani2020,Basar2019}}.
	An RSS is composed of many passive reflecting elements (REs) 
	{forming} a planar surface {that interacts in a controllable manner with electromagnetic waves.} The RSS controls the wireless propagation environment by 
	{smartly adjusting the amplitude and phase} of each RE through the controller. The 
	contribution of all REsleads to a beamforming-like behaviour, where incident radio signals are ``reflected'' to enhance the 
	signal power at the receiver. 
	{Due to the nearly passive nature of RSS and the absence of radio frequency chains, RSS are more energy-efficient and cost-effective than conventional relays.}
	
	Inspired by the 
	{potentials} of RSS, several works 
	{investigated the phase shift design for RSS-assisted communications.}
	For instance, {authors of \cite{wu2019intelligent} jointly optimized the transmit beamforming of a multi-antenna transmitter and the RSS configuration, aiming to maximize the  achievable data rate at one receiver. Results indicated the RSS-assisted communication achieved a comparable performance to conventional relay-assisted systems.} 
	Yang \textit{et al.} proposed in \cite{yang2020intelligent} an RSS-asssited communication, where transmissions follow the orthogonal frequency division multiplexing (OFDM) protocol. Channel correlation of OFDM transmissions has been leveraged to reduce estimation overhead. Then, transmit power and RSS configuration have been optimized for data rate maximization, similarly to \cite{wu2019intelligent}. 
	Given a similar system model, authors of \cite{ZhengZhang-2019} 
	{designed an RSS phase-shift pattern that minimizes channel estimation errors. Subsequently, RSS phase shifts have been optimized by relying on the strongest tap of the channel gain.}
	
	{Nevertheless, the aforementioned works considered either perfect channel state information (CSI) \cite{wu2019intelligent}, or continuous phase shifts for RSS configuration \cite{yang2020intelligent,ZhengZhang-2019}. 
		{Indeed, channel} estimation is a major challenge and perfect CSI is practically difficult to realize. Also, continuous phase shifting requires RSS with infinite resolution shifters, which is still infeasible. Since RSSs have typically many REs, it is impractical to consider high resolution quantized phase shifters. 
		Recently, a few studies investigated RSS-assisted communications with quantized phase shifting. In \cite{zhang2020reconfigurable}, the impact of a limited number of phase shifts on the data rate is investigated, while authors of \cite{Changsheng2019} proposed an algorithm to optimize the quantized RSS configuration to achieve a maximized data
		rate.}

	{Motivated by the aforementioned limitations, we tackle in this paper the problem of maximizing the data rate of 
		an RSS-assisted communication that leverages the OFDM transmission protocol, {and} under quantized phase shifts and channel estimation errors consideration. 
		{First, we rely on pilot signals with given predefined RSS configurations, to estimate the overall transmitter-RSS-receiver channel using least-square error (LSE) channel estimator.}
		Then, we optimize the quantized phase shifting-based RSS configuration. Specifically, we propose a low-complex greedy approach that maximizes the data rate at the receiver.
		Through simulations, we demonstrate that our algorithm outperforms existing benchmarks. 
		{Moreover, we evaluate the impact of several factors on the data rate. These factors include the type of communication channels (line-of-sight (LoS) or non-LoS), quantization resolution, and RSS placement.}
	}
	
	
	
	
	The remaining of the paper is organized as follows. The system model is presented
	{in Section \ref{sec:SystemDescription}. 
		Section \ref{sec:ProblemFormulation} formulates the problem and discusses conventional solutions. 
		The proposed approach is detailed in Section \ref{sec:proposed_solution}.}
	Section \ref{sec:Simulation_results} shows the simulation results, and finally Section \ref{sec:conclusion} closes the paper.
	
	
	

	{\section{System Model} \label{sec:SystemDescription}
	}
	
	\subsection{Transmission Model}

	{We consider the downlink of a wireless system, composed of an access point ($s$) transmitting to a user equipment ($d$), directly or through 
		wall-mounted RSS ($r$).
	} We assume OFDM transmissions and pilot based channel estimation where signals are reflected with a predefined set of RSS configurations. 
	%
	The RSS has $N$ REs connected to a controller that dynamically adjusts phase shifting by changing REs' states. Indeed, each RE can be set to one of the available $L=2^b$ states, where state $s_i=(b_i, \theta_i)$ is characterized by an amplitude loss $b_i \in [0,1]$ and a phase shift $\theta_i \in [0, 2 \pi)$, $\forall i=1,\ldots,L$, and $b$ is the number of states quantization bits.

	
	
	For simplicity's sake, we assume static wireless channels $s-r$, $r-d$, and $s-d$ during one transmission, but variable among different transmissions.
	Based on the OFDM protocol, we assume that $K$ sub-carriers are used to transmit with equal powers $P/K$ and equal bandwidths $B/K$, where $P$ is the transmit power of $s$, and $B$ is the available bandwidth.
	
	
	
	At $s$, each OFDM symbol $\mathbf{\tilde x} = [x_1,\dots,x_{K}]$, where $\mathop{\mathbb{E}}(x_k^2)=P/K$, is transformed using the K-point inverse discrete Fourier transform (IDFT). Then, a guard interval of length $L_{g}$ is inserted. We assume that $L_{g}\geq G$, where $G$ is the number of taps of the baseband finite impulse response (FIR) filter representing the channel. At $d$, the guard interval is removed and a K-point DFT is performed. Then, the baseband received signal is obtained in the frequency domain as \cite{ZhengZhang-2019}
	
	\vspace{-11pt}
	\begin{equation}
	\small
	\label{eq:Rx_signal}
	\mathbf{\tilde z} = \mathbf{\tilde X}\, \Big(\sum_{n = 1}^N \mathbf{\tilde a}_n\, \omega_{n}\, \odot\, \mathbf{\tilde b}_n + \fr h_{sd}\Big) + \mathbf{\tilde n}  \in \mathbb{C}^{K\times 1},
	\end{equation}
	
	\vspace{-10pt}
	\noindent
	where $\fr X = \text{diag}(\fr x) \in \mathbb{C}^{K\times K}$, $\fr a_n \in \mathbb{C}^{K\times 1}$ is the channel frequency response (CFR) of the link between the transmitter and the RSS at the $n^{th}$ RE, $\fr b_n \in \mathbb{C}^{K\times 1}$ is the CFR of the link between the RSS and the receiver for the $n^{th}$ RE, $\fr h_{sd}$ is the CFR of the direct link, $\omega_{n}=\beta_n e^{j \theta_n}$ is the reflection coefficient of the $n^{th}$ RE, and $\fr n$ is the additive white Gaussian noise (AWGN) vector whose entries are distributed as $\mathcal{CN}(0,\, \sigma^2)$. Finally, $\odot$ denotes the Hadamard product.
	Then, the summation term in (\ref{eq:Rx_signal}) is rewritten as \cite[eq. (24)]{bjornson2021reconfigurable}
	
	\vspace{-12pt}
	\begin{equation}
	\small
	\sum_{n = 1}^N \mathbf{\tilde a}_n\, \omega_{n}\, \odot\, \mathbf{\tilde b}_n = \mathbf{F} \mathbf{V}^T \bm{\omega} = \mathbf{F} \mathbf{h}_{srd} = \fr h_{srd},
	\end{equation}
	
	\vspace{-10pt}
	\noindent
	where $\mathbf{V} \in \mathbb{C}^{N \times G}$ is the cascaded channel in the time domain, 
	$\bm{\omega} = [\omega_{1},\,\dots,\,\omega_{N}]^T$, and $\mathbf{F}$ is the $K \times G$ DFT matrix having $(\nu, k)^{th}$ element as $e^{-j 2 k \nu / K}$. Finally, $(\cdot)^T$ is the transpose operator.  Subsequently, \eqref{eq:Rx_signal} can be given as
	
	\vspace{-10pt}
	\begin{equation}
	\small
	\label{eq:Rx_signal2}
	\mathbf{\tilde z} =\fr{X} \left(\fr h_{srd} + \fr h_{sd}\right) + \fr n = \fr X \fr h + \fr n,
	\end{equation}
	
	\vspace{-10pt}
	\noindent
	where $\fr h_{srd}$ is the CFR of overall $s-r-d$ channel, $\fr h_{sd}$ is the CFR of channel $s-d$ and $\fr h$ stands for their summation. 
	For the remaining, we assume an obstructed direct link ($\fr {h}_{sd}=0$).
	
	\subsection{Channel Estimation}
	Several approaches have been proposed to estimate the channel 
	{of} RSS-assisted communications. For instance, it is estimated in \cite{Changsheng2019} using training with the designed reflection patterns, while authors of \cite{ZhengZhang-2019} proposed a DFT reflection pattern design that builds an orthogonal matrix with each entry under the unit-modulus constraint. Also, authors of \cite{bjornson_pilotnumber} proposed pilot signals estimation, where different RSS configurations are set up for the pilot signals transmission.


	Similarly to \cite{bjornson2021reconfigurable}, we assume here that channel estimation is based on the transmission of $N$ pilot signals, each with a different RSS configuration $\bm{\omega_{\theta_i}}$, $\forall i=1,\ldots,N$. 
	First, we define the minimal number of pilot sub-carriers as $T=\ceil{\pi N A / \lambda^2}-1$, where $A$ is the area of each RE, $\lambda$ is the wavelength, and $\ceil{\cdot}$ is the ceiling function \cite{bjornson_pilotnumber}. 
	The joint received signal can be written as 
	
	\vspace{-10pt}
	\begin{equation}
	\small
	\fr Z =\fr{X} \mathbf{F} \mathbf{V}^T \mathbf{\Omega} + \fr N \; \in \mathbb{C}^{K \times N},
	\end{equation}
	
	\vspace{-10pt}
	\noindent
	where $\fr Z = [\fr z_1,\, \dots,\, \fr z_N]$, $\fr z_i$ is defined as in (\ref{eq:Rx_signal2}), $ \mathbf{\Omega} = [\bm{\omega_{\theta_1}},\,\dots,\,\bm{\omega_{\theta_N}}] \in \mathbb{C}^{N \times N}$, 
	$\fr N =  [\fr n_1,\, \dots,\, \fr n_N] \in \mathbb{C}^{N \times N}$, and $\fr n_i$ is defined as $\fr n$, $\forall i=1,\ldots,N$. 
	For the sake of simplicity, we 
	{adopt} the LSE channel estimation for the RSS-assisted link as in \cite{bjornson2021reconfigurable}. 
	Hence, the estimated $s-r-d$ channel is
	
	\vspace{-10pt}
	\begin{equation}
	\label{estimate}
	\small
	\fr{\mathbf{V}}^T = \mathbf{F}^{-1} \fr X^{-1} \fr Z \; \mathbf{\Omega}^{-1}.
	\end{equation}

	\vspace{-15pt}

	\subsection{Achievable Data Rate}
	
	Assuming that the channel is estimated as in (\ref{estimate}), then, the achievable data rate for imperfect CSI (iCSI) is calculated as

	\vspace{-18pt}
	\begin{equation}
	\small
	\label{eq:achievable-rate2}
	R^{\rm iCSI} = \frac{B}{K + G - 1}\sum_{k = 1}^{K} \log_2 \left(1+ \frac{P \left\lvert \mathbf{f}_k^H \fr{\mathbf{V}}^T \boldsymbol{\omega} \right \rvert^2 }{K B \sigma^2} \right),
	\end{equation}
	
	\vspace{-8pt}
	\noindent
	where $\textbf{f}_k^H$ is the $k^{th}$ row of the DFT matrix $\textbf{F}$. Eq.
	(\ref{eq:achievable-rate2}) expresses the data rate as a summation over $K$ sub-carriers, divided by $K+G-1$ to make up for the guard interval loss.

	

	
	
	
	
	\vspace{-10pt}
	\section{Problem Formulation} \label{sec:ProblemFormulation}
	
	\vspace{-5pt}
	In this paper, we aim to maximize the receiver's data rate for the RSS-assisted communication, under the quantized phase shifting constraint. {For the sake of simplicity, we consider lossless reflectors, i.e., $\beta_n=1$, $\forall n=1,\ldots,N$.} We formulate the problem as follows:

	\vspace{-15pt}
	\begin{subequations}
		\small
		\begin{align}
		\label{eq:Capacity}
		\max_{\boldsymbol{\omega}} \quad & R^{\rm iCSI} \tag{P1} \\
		\label{c1} \textrm{s.t.} \quad & \theta_n \in \left\{ \delta_1, \ldots \delta_L \right\}, \forall n=1,\ldots,N, \tag{P1.a}
		\end{align}
	\end{subequations}
	
	\vspace{-7pt}
	\noindent
	where $\delta_l \in [0,2 \pi)$ is the $l^{th}$ quantized phase-shift level, $\forall l=1,\ldots,L$, and $\delta_1< \delta_2< \ldots < \delta_L$. This problem is equivalent to maximize the energy over all OFDM sub-carriers at the receiver. The latter has been proven non-convex, and hence a tractable global solution cannot be obtained \cite{bjornson2021reconfigurable}. Nevertheless, several heuristic solutions have been proposed in the literature. For instance, authors of \cite{bjornson2021reconfigurable} introduced the strongest tap maximization (STM) algorithm, which seeks the RSS configuration that improves the signal quality over one channel tap. Then, it selects the RSS configuration that provided the highest amplitude received signal. It has been shown that this type of approaches is useful when the communication channel is dominated by a LoS component, which is stronger than all other taps. Alternatively, authors of \cite{ZhengZhang-2019} proposed a method that maximizes the sum of channel power gain (SCPGM). It relies on the strongest channel tap. 
	
	Both STM and SCPGM perform near optimality when channels are dominated by one tap, as in the case of LoS. However, these methods are strongly handicapped for non-LoS receivers. In addition, STM and SCGPM solutions were intitially developed 
	for a continuous phase shift range, but can be mapped --with some performance degradation-- to quantized phase shifts. STM and SCPGM will be used later (in Section V) as benchmarks for comparison with our proposed solution. The latter is detailed in the next section.

	\vspace{-8pt}
	\section{Proposed Solution} 
	\vspace{-5pt}
	\label{sec:proposed_solution}
	Our goal is to propose a greedy approach for RSS configuration that maximizes $R^{\text{iCSI}}$, under the quantized phase shift constraint. Given an initial RSS configuration, the key idea lies in iteratively evaluating the effect of changing the phase shift of one RE at a time, by calculating the associated data rate, using (\ref{eq:achievable-rate2}), at each iteration. If performance is improved, then the REs' phase shifts are updated with new ones. 
	
	Specifically, let $\boldsymbol{\omega}_0^i$ be the initial RSS configuration vector in iteration $i$. Initialization can be random or based on existing solutions such as STM or SCPGM. We denote also by $\mathcal{W}_r^i=\{\boldsymbol{\omega}_{1,r}^i, \ldots,\boldsymbol{\omega}_{L-1,r}^i\}$ the set of possible RSS configurations when changing the phase shift value of the $r^{th}$ RE, while keeping the remaining REs unchanged, with $\boldsymbol{\omega}_{l,r}^i$ is the $l^{th}$ RSS configuration in $\mathcal{W}_r^i$, corresponding to the change of RE $r$ phase shift to the $l^{th}$ quantized value, $\forall r=1,\ldots,N$, $l-1,\ldots,L-1$.  
	For each $\mathcal{W}_r^i$, we evaluate the corresponding data rates using (\ref{eq:achievable-rate2}), and store them as $\mathcal{R}_r^i=\{R_{1,r}^i,\ldots,R_{L-1,r}^i \}$. Then, we create set $\bar{\mathcal{R}}^i=\{ \max \mathcal{R}_r^i \; | \; \max \mathcal{R}_r^i >R_0^i, \; r=1,\ldots,N \}$, where $R_0^i$ is the performance of the initial configuration $\boldsymbol{\omega}_0^i$, and we define the corresponding configurations' set $\bar{\mathcal{W}}^i$. 
	
	Assuming that $\bar{\mathcal{R}}^i$ is ordered from the highest to the lowest value, then, given parameter $\alpha \in (0,1]$, we look into the REs phase shift values in only the first $\alpha W_i$ RSS configurations, where $W_i$ is the size of the set $\bar{\mathcal{W}}^i$, and create $\boldsymbol{\omega}_0^{i+1}$ from $\boldsymbol{\omega}_0^i$ by setting the REs' phase shifts to the changed ones from the $\alpha W_i$ first configurations of $\bar{\mathcal{W}}^i$. 
	The previous steps are repeated until $\bar{\mathcal{R}}^I=\emptyset$, where $I$ is the last iteration, which corresponds to the algorithm stop.
	The operation of our approach is described in Algo. \ref{Algo00}.
	
	\begin{algorithm}[h]
		\small{
			\caption{Proposed Algorithm}
			\label{Algo00}
			\begin{algorithmic}[1]
				\State Initialize RSS configuration $\boldsymbol{\omega}_0^0$, data rate $R_0^0$, and $\bar{\mathcal{R}}^0=\{ R_0^0 \}$
				\State Set $i=1$ and $\alpha \in (0,1]$
				\While {$\bar{\mathcal{R}}^{i-1} \neq \emptyset$} \do \\
				\State Build $\mathcal{W}_r^{i}$ and $\mathcal{R}_r^i$, $\forall r=1,\ldots, N$
				\State Create $\bar{\mathcal{R}}^{i}$ and order it from highest to lowest value
				\State Identify the REs' phase shift changed values (with respect to $\boldsymbol{\omega}_0^{i-1}$) in the first $\alpha W_i$ RSS configurations of $\bar{\mathcal{W}}^i$ and inject them into $\boldsymbol{\omega}_0^{i-1}$ to create $\boldsymbol{\omega}_0^{i}$
				\State Update $i=i+1$
				\EndWhile
				\State Return $\boldsymbol{\omega}_0^{i-1}$ 
		\end{algorithmic}}
	\end{algorithm}

	To further enhance the performance of our method, we propose to parallelize several replicas of our algorithm, but with different initializations, e.g., random, STM, and SCPGM. Then, the replica that presents the best performance is selected. 
	we call this approach Algo. 2. 

	\vspace{-7pt}
	\section{Simulation Results} \label{sec:Simulation_results}
	\vspace{-5pt}
	To evaluate the performance of the proposed algorithm, an OFDM transmission with a bandwidth $B=10$ MHz is simulated. We assume that $K=500$ OFDM sub-carriers and $G=20$ taps for each channel. 
	Channels follow the Rician fading model, where LoS links are assumed having K-factor $\kappa = 10$ dB, while non-LoS (NLoS) links have $\kappa = -10$ dB \cite{yildirim2020modeling}. Unless stated otherwise, we assume  the distances between $s-r$ and $r-d$ are fixed to $d_{sr}=100$ m and $d_{rd}=15$ m, respectively, and are prone to path loss effect with exponents $\beta=2$ and $\beta=2.75$, respectively. Moreover, noise variance $\sigma^2=-165.14$ dBm/Hz and the number of REs of the RSS is fixed to $N=4096$. Without loss of generality, we assume that the $s-r$ link is always in LoS, while the $r-d$ link can be either LoS or NLoS. Finally, given $L$ quantization levels, we assign phase shift values from the set $\left\{ 0, \frac{2 \pi}{L}, \frac{4 \pi }{L} \ldots, \frac{2 \pi (L-1)}{L} \right\}$.\footnote{The simulation results can be reproduced using code available in \cite{Git}.} The following results are depicted as functions of averaged signal-to-noise ratio (SNR), given by
	
	\vspace{-8pt}
	\begin{equation}
	\small 
	\label{eq:SNR}
	\text{SNR} = \frac{P\, d_{sr}^{-\beta_{sr}}d_{rd}^{-\beta_{rd}}}{K\,B\,\sigma^2},
	\end{equation}
	
	\vspace{-6pt}
	\noindent
	where $\beta_{sr}$ and $\beta_{rd}$ are the path loss exponents. 

	\begin{figure*}[t]
		\begin{minipage}{0.3\linewidth}
			\includegraphics[width=160pt, height=110pt]{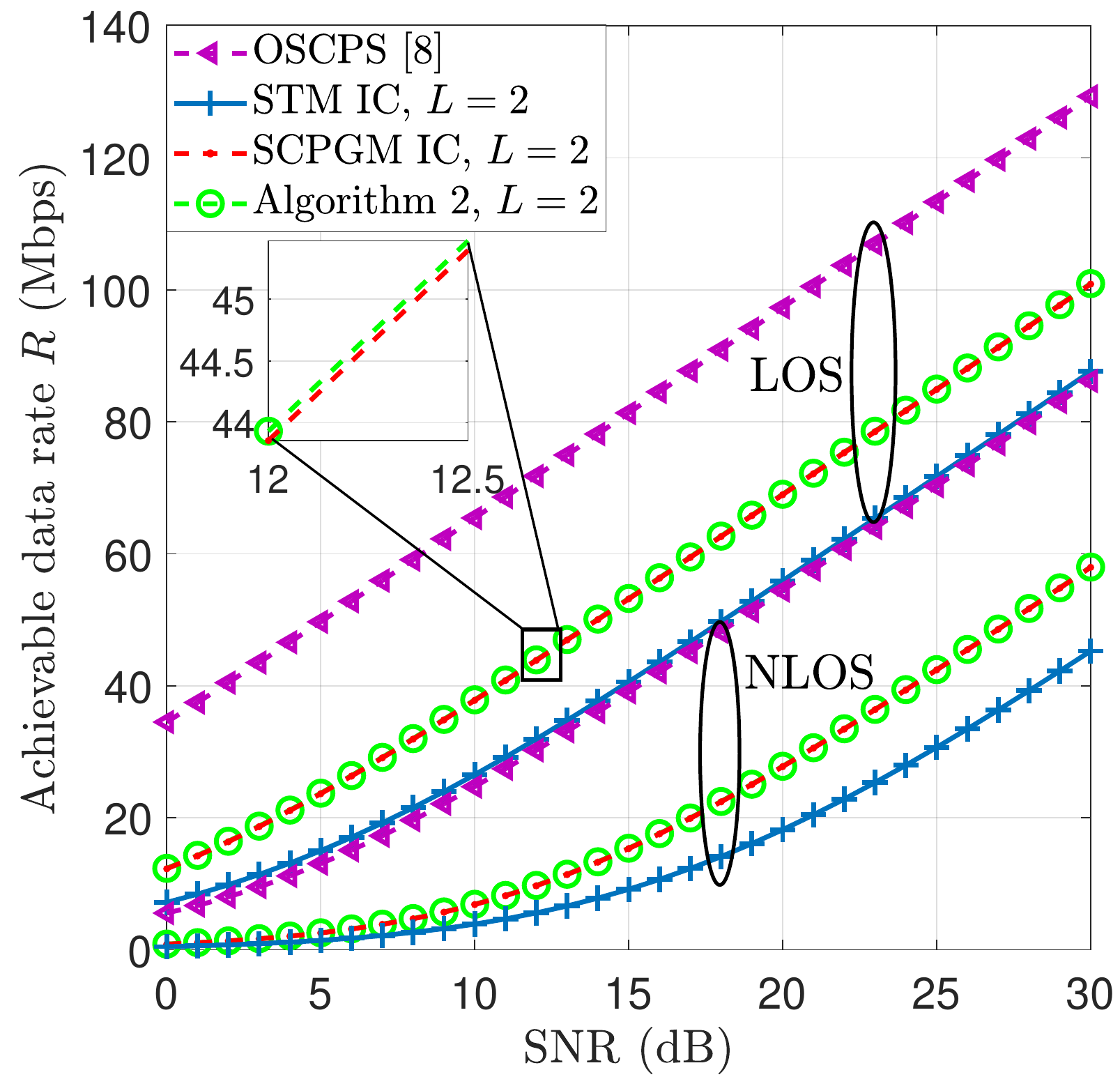}
			\caption{Achievable data rate vs. SNR ($L=2$).}
			\label{fig:Data_rate_SNR_b1_comparison}
		\end{minipage}
		\hfill
		\begin{minipage}{0.3\linewidth}
			\includegraphics[width=155pt,height=105pt]{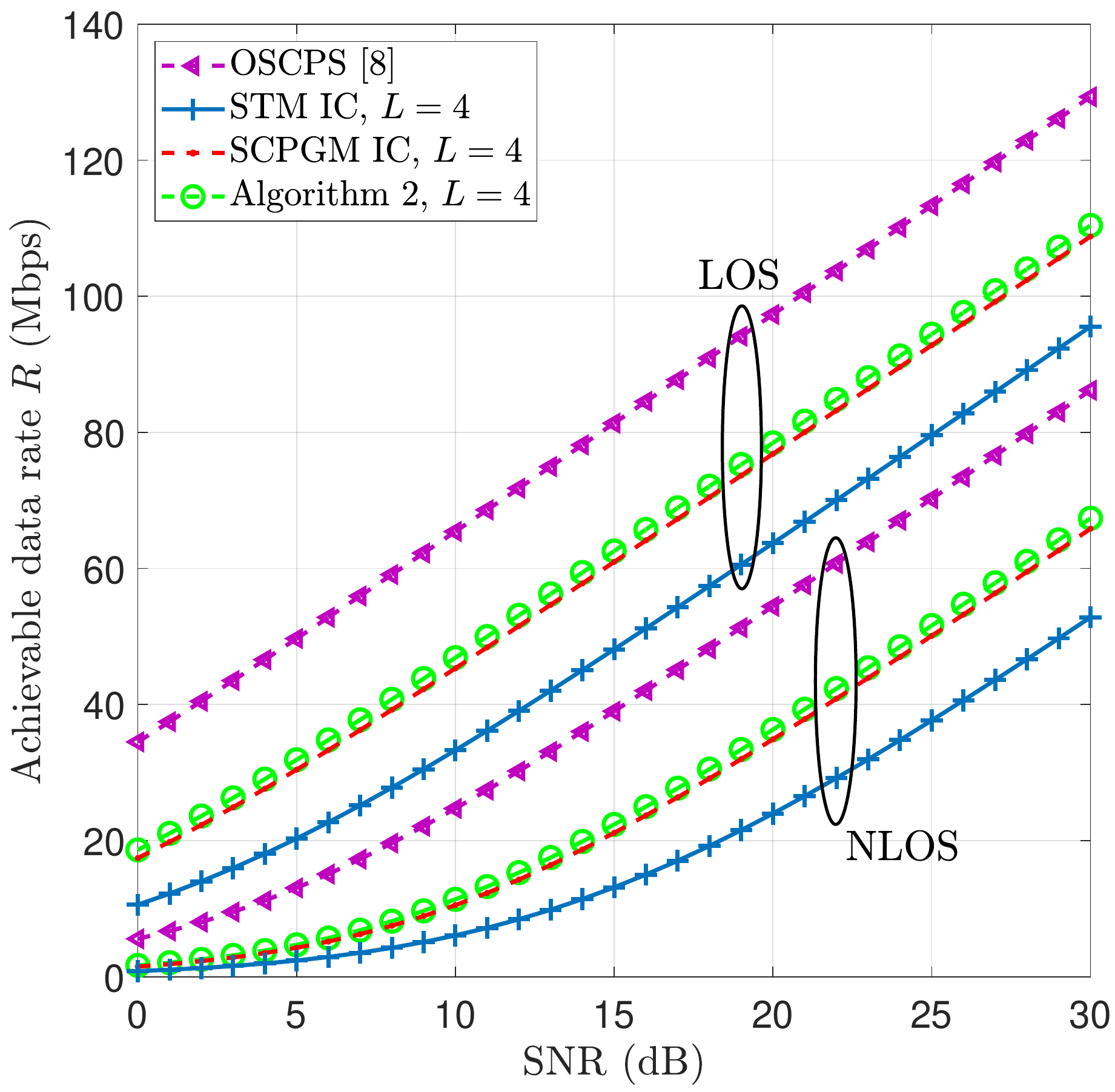}
			\caption{Achievable data rate vs. SNR ($L=4$).}
			\label{fig:Data_rate_SNR_b2_comparison}
		\end{minipage}
		\hfill
		\begin{minipage}{0.3\linewidth}
			\includegraphics[width=147pt,height=100pt]{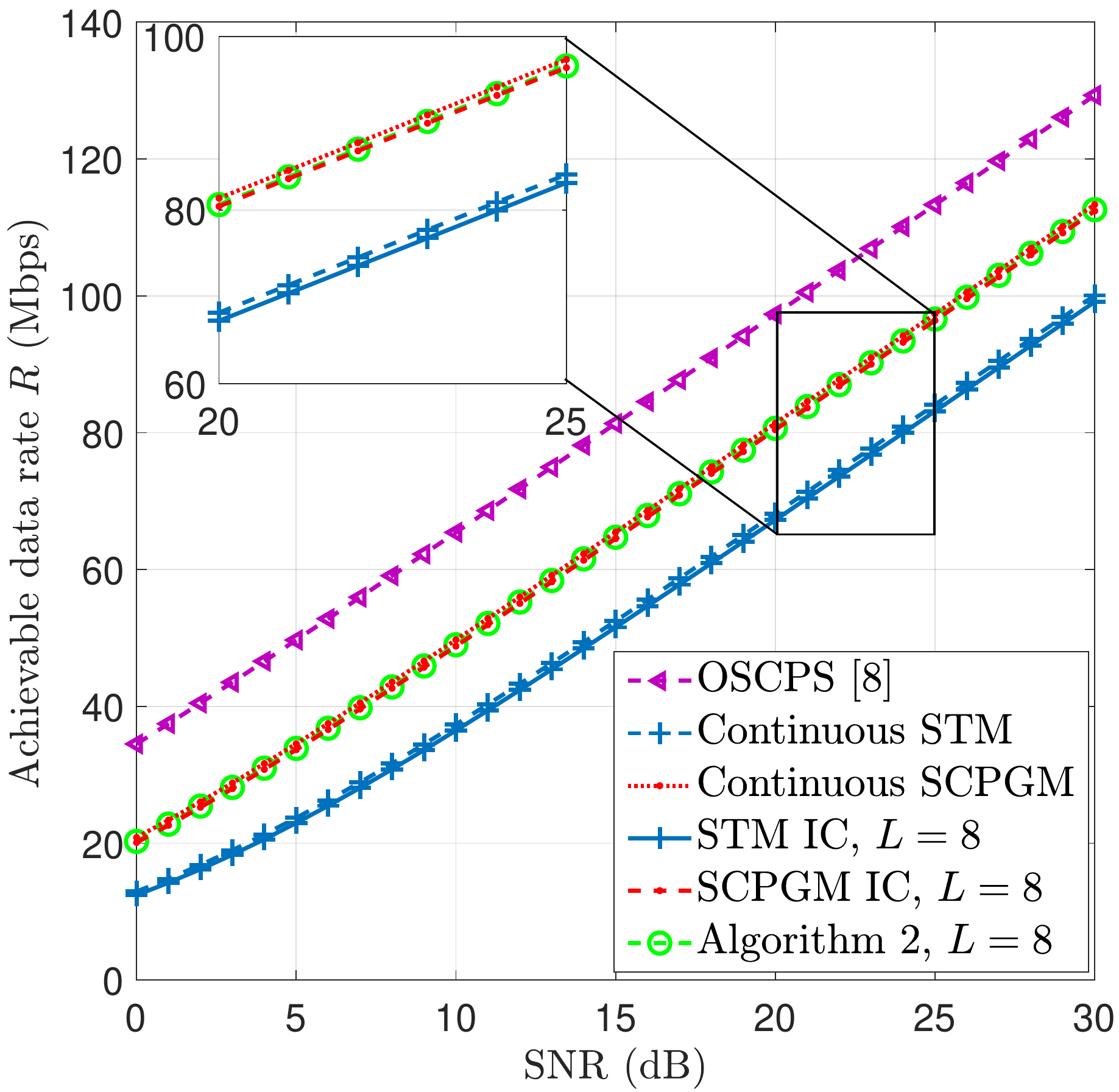}
			\caption{Achievable data rate vs. SNR ($L = 8$, LoS).}
			\label{fig:continuous}
		\end{minipage}
	\end{figure*}

	\begin{figure*}[t]
		\begin{minipage}{0.3\linewidth}
			\includegraphics[width=155pt,height=100pt]{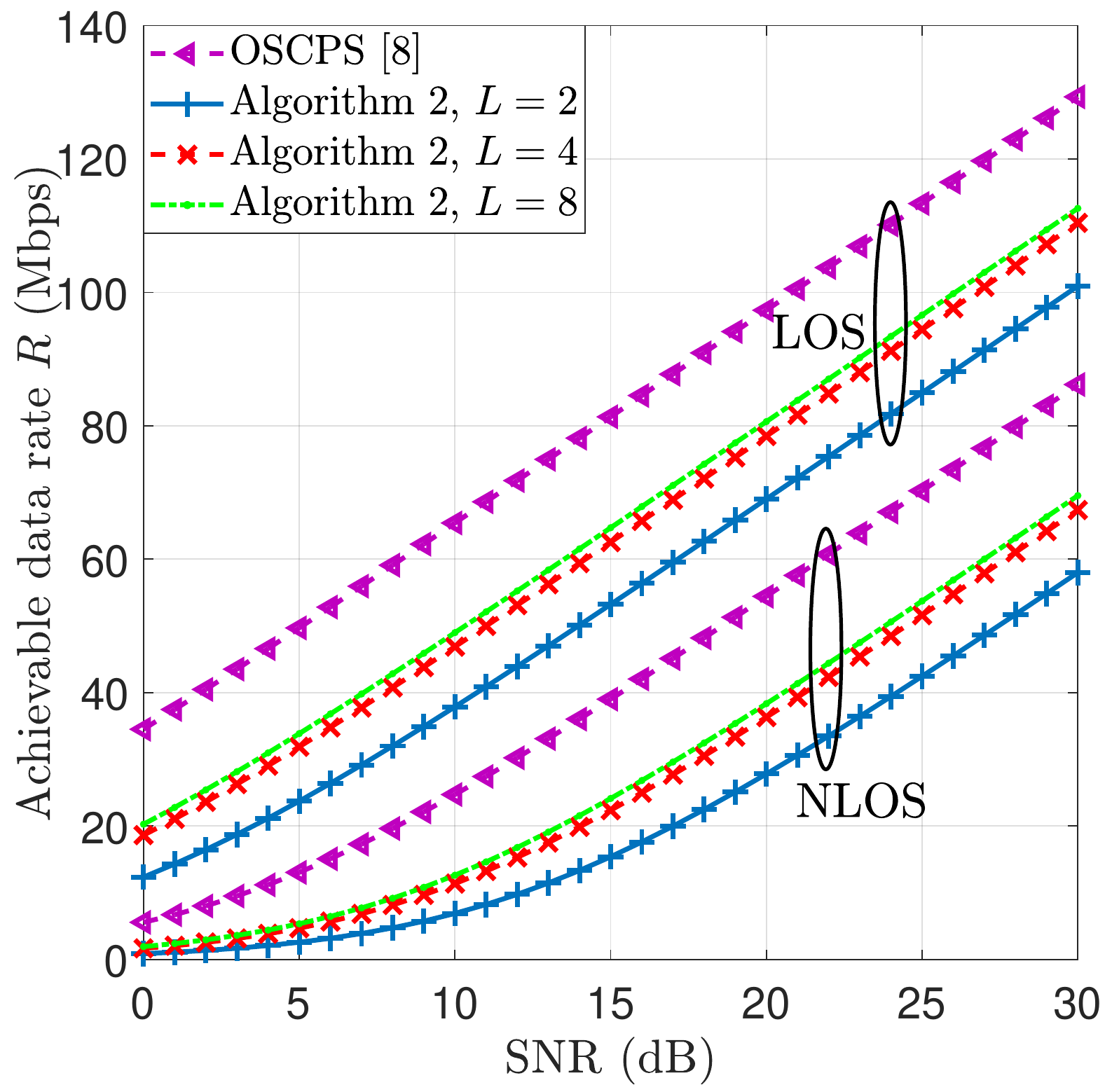}
			\caption{Achievable data rate vs. SNR (different quantization levels).}
			\label{fig:Data_rate_SNR_bits_comparison}
		\end{minipage}
		\hfill
		\begin{minipage}{0.3\linewidth}
			\includegraphics[width=160pt,height=105pt]{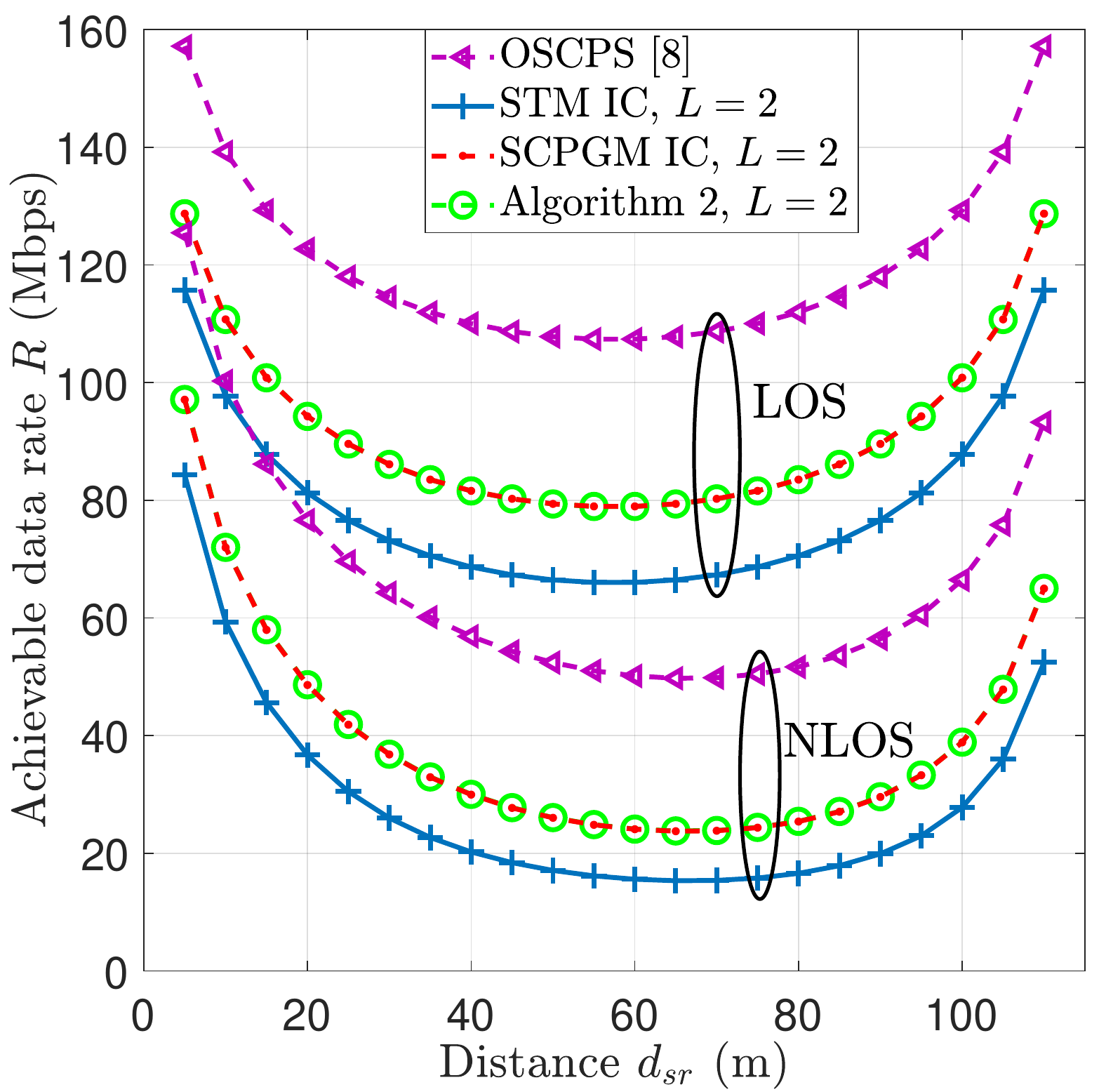}
			\caption{Achievable data rate vs $d_{sr}$.}
			\label{fig:Data_rate_SNR30_b1_Distance}
		\end{minipage}
		\hfill
		\begin{minipage}{0.3\linewidth}
			\includegraphics[width=145pt,height=105pt]{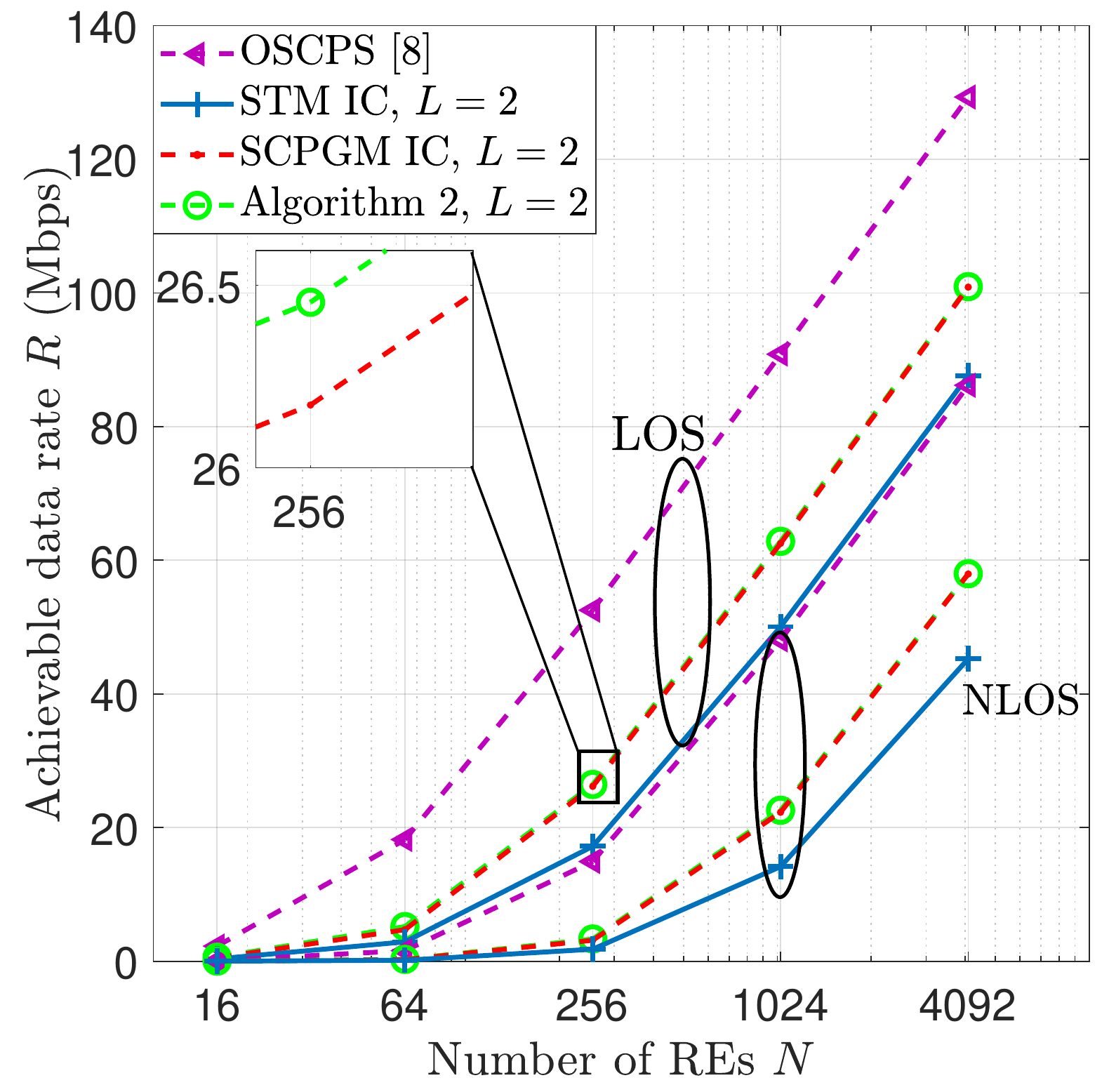}
			\caption{Achievable data rate vs nbr. of REs $N$.}
			\label{fig:Data_rate_SNR30_b1_N}
		\end{minipage}
	\end{figure*}


	
	Fig. \ref{fig:Data_rate_SNR_b1_comparison} presents the achievable data rate as a function of SNR  given LoS and NLoS $r-d$ links respectively, and with $L=2$. It compares between Algo. 1 initialized with STM solution (called STM IC), Algo. 1 initialized with SCPGM solution (denoted SCPGM IC), proposed Algo. 2\footnote{In the remaining, $\alpha=1$ for Algo. 1. This is motivated by the faster convergence and better results compared to cases with smaller $\alpha$.}, and a benchmark OSCPS (Optimal Solution using Continuous Phase Shift, proposed in \cite[eq. (57)]{bjornson2021reconfigurable}).
	First, for any channel condition, 
	the data rate improves proportionally to SNR. Also, Algo. 2 outperforms STM IC and slightly SCGPM IC, but is under-performing compared to OSCPS. Indeed, due to the multiple parallel initializations, Algo. 2 is capable of exploring other RSS configuration solutions that enhance the performance. However, due to quantization errors, its gap with OSCPS is still significant, around 20 Mbps (i.e., 2 bps/Hz). Finally, the system with LoS $r-d$ link achieves up to 200\% data rate performance of the system having a NLoS $r-d$ link. 
	

	
	Similar results are depicted in Fig. \ref{fig:Data_rate_SNR_b2_comparison} for $L=4$. Here, we notice that Algo. 2 is superior to both STM IC and SCPGM IC, with a larger performance gap than in the previous figure. 
	Indeed, with a higher $L$, Algo. 2 explores a larger size solution space than for $L=2$, which allows to obtain better results. Finally, we notice that having $L=4$ improves the data rate for Algo. 2 compared to the case of $L=2$.  
	
	In Fig. \ref{fig:continuous}, we extend the comparison to $L=8$, LoS, and to continuous STM/SCPGM benchmarks where phase shifts are not quantized. Algo. 2 outperforms continuous STM and is similar to continuous SCGPM. Also, STM IC and SCPGM IC present close performances to their continuous benchmarks. This proves the efficiency of our algorithm in achieving high performance despite the quantization information loss.
	
	In Fig. \ref{fig:Data_rate_SNR_bits_comparison}, the impact of different quantization levels is evaluated for Algo. 2 and OSCPS benchmark. As it can be seen, increasing $L$ improves the data rate for any $r-d$ channel condition. However, the gain gap between $L=4$ and $L=8$ is very small compared to the one between $L=2$ and $L=4$. Thus, we identify a trade-off between the system's complexity, due to the number of phase shift quantization levels, and the data rate performance. Practically, one would carefully select the number of quantization levels to avoid complexifying the system for an insignificant performance gain.

	

	Fig. \ref{fig:Data_rate_SNR30_b1_Distance} presents the achievable data rate as a function of the distance $d_{rd}$ ($L = 2$ and SNR$= 30$ dB). Assuming that $d_{sr}+d_{rd}=115$ m, we move the RSS either closer or further from the transmitter. As it can be seen, placing the RSS near the transmitter or the receiver in LoS conditions is equally preferable. In contrast, when the $r-d$ link is in NLoS, placing the RSS closer to the transmitter $s$ is recommended.

	Fig. \ref{fig:Data_rate_SNR30_b1_N} illustrates the data rate as a function of the number of REs $N$, for $L = 2$ and SNR$= 30$ dB. As $N$ increases, the data rate improves for all algorithms. However, such an improvement comes at the expense of a longer processing time of the RSS configuration algorithms.

	\vspace{-8pt}
	\section{Conclusion} \label{sec:conclusion}
	\vspace{-5pt}
	In this paper, we proposed a novel greedy configuration algorithm for RSS-assisted communication systems. We proved that proposed Algo. 2 outperforms benchmarks. Also, the impact of several parameters was explored, which provided novel guidelines: 1) practical RSS phase shift quantization requires a low number of states, e.g., four quantization states achieve acceptable data rates, and eight states achieve almost the same performance as the continuous SCPGM, 2) it is recommended to place the RSS either closer to transmitter or receiver in LoS conditions, but only closer to transmitter for NLoS link towards the receiver or vice-versa, and 3) it is preferable to deploy RSS with the highest number of REs.
	As a future work, the selection of quantization phase shift values will be optimized to enhance the communication's performance in different propagation environment conditions.
	
	\vspace{-5pt}
	
	\bibliographystyle{IEEEtran}
	\bibliography{IEEEabrv,tau2021}

\begin{thebibliography}{10}
\providecommand{\url}[1]{#1}
\csname url@samestyle\endcsname
\providecommand{\newblock}{\relax}
\providecommand{\bibinfo}[2]{#2}
\providecommand{\BIBentrySTDinterwordspacing}{\spaceskip=0pt\relax}
\providecommand{\BIBentryALTinterwordstretchfactor}{4}
\providecommand{\BIBentryALTinterwordspacing}{\spaceskip=\fontdimen2\font plus
\BIBentryALTinterwordstretchfactor\fontdimen3\font minus
  \fontdimen4\font\relax}
\providecommand{\BIBforeignlanguage}[2]{{%
\expandafter\ifx\csname l@#1\endcsname\relax
\typeout{** WARNING: IEEEtran.bst: No hyphenation pattern has been}%
\typeout{** loaded for the language `#1'. Using the pattern for}%
\typeout{** the default language instead.}%
\else
\language=\csname l@#1\endcsname
\fi
#2}}
\providecommand{\BIBdecl}{\relax}
\BIBdecl

\bibitem{Alfattani2020}
S.~Alfattani, W.~Jaafar, Y.~Hmamouche, H.~Yanikomeroglu, A.~Yongaçoglu, N.~D.
  Dao, and P.~Zhu, ``Aerial platforms with reconfigurable smart surfaces for
  {5G} and beyond,'' \emph{IEEE Commun. Mag.}, vol.~59, no.~1, pp. 96--102,
  Jan. 2021.

\bibitem{Basar2019}
E.~{Basar}, M.~{Di Renzo}, J.~{De Rosny}, M.~{Debbah}, M.~{Alouini}, and
  R.~{Zhang}, ``Wireless communications through reconfigurable intelligent
  surfaces,'' \emph{IEEE Access}, vol.~7, pp. 116\,753--116\,773, Aug. 2019.

\bibitem{wu2019intelligent}
Q.~Wu and R.~Zhang, ``Intelligent reflecting surface enhanced wireless network
  via joint active and passive beamforming,'' \emph{IEEE Trans. Wireless
  Commun.}, vol.~18, no.~11, pp. 5394--5409, Nov. 2019.

\bibitem{yang2020intelligent}
Y.~Yang, B.~Zheng, S.~Zhang, and R.~Zhang, ``Intelligent reflecting surface
  meets {OFDM}: Protocol design and rate maximization,'' \emph{IEEE Trans.
  Commun.}, vol.~68, no.~7, pp. 4522--4535, Jul. 2020.

\bibitem{ZhengZhang-2019}
B.~Zheng and R.~Zhang, ``Intelligent reflecting surface-enhanced {OFDM:}
  channel estimation and reflection optimization,'' \emph{IEEE Wireless Commun.
  Lett.}, vol.~9, Apr. 2020.

\bibitem{zhang2020reconfigurable}
H.~Zhang, B.~Di, L.~Song, and Z.~Han, ``Reconfigurable intelligent surfaces
  assisted communications with limited phase shifts: How many phase shifts are
  enough?'' \emph{IEEE Trans. Veh. Technol.}, vol.~69, no.~4, pp. 4498--4502,
  Apr. 2020.

\bibitem{Changsheng2019}
C.~You, B.~Zheng, and R.~Zhang, ``Intelligent reflecting surface with discrete
  phase shifts: Channel estimation and passive beamforming,'' in \emph{Proc.
  IEEE Int. Conf. Commun. (ICC)}, 2020, pp. 1--6.

\bibitem{bjornson2021reconfigurable}
E.~Björnson, H.~Wymeersch, B.~Matthiesen, P.~Popovski, L.~Sanguinetti, and
  E.~de~Carvalho, ``Reconfigurable intelligent surfaces: A signal processing
  perspective with wireless applications,'' \emph{ArXiv}, 2021.

\bibitem{bjornson_pilotnumber}
E.~Björnson and L.~Sanguinetti, ``Rayleigh fading modeling and channel
  hardening for reconfigurable intelligent surfaces,'' \emph{IEEE Wireless
  Commun. Lett.}, vol.~10, no.~4, pp. 830--834, Apr. 2021.

\bibitem{yildirim2020modeling}
I.~Yildirim, A.~Uyrus, and E.~Basar, ``Modeling and analysis of reconfigurable
  intelligent surfaces for indoor and outdoor applications in future wireless
  networks,'' \emph{IEEE Trans. Commun.}, vol.~69, no.~2, pp. 1290--1301, Feb.
  2021.

\bibitem{Git}
M.~Rivera and M.~Chegini, ``{RIS Phase Shift Optimizer},''
  \url{https://github.com/moh-C/RISPhaseShiftOptimizer}, [Online; accessed
  30-Aug-2021].

\end{thebibliography}

\end{document}